\documentclass[reprint,superscriptaddress,amsmath,amssymb,aps,prb,floatfix,]{revtex4-2}
\usepackage{graphicx}

\usepackage{bm}
\usepackage{relsize}
\usepackage{amsmath}
\usepackage{amsfonts}
\usepackage{mathtools}
\usepackage{braket}
\usepackage{url}
\usepackage[colorlinks = true,
            linkcolor = blue,
            urlcolor  = blue,
            citecolor = blue,
            anchorcolor = blue]{hyperref}
\usepackage{footmisc}
\usepackage[utf8]{inputenc}
\usepackage[english]{babel}
\usepackage[dvipsnames]{xcolor}
\usepackage[T1]{fontenc}
\usepackage{comment}
\begin{document}

\title{Anomalous dispersion via dissipative coupling \\ in  a quantum well exciton-polariton microcavity
}

\author{D.~Biega\'{n}ska}
\email{dabrowka.bieganska@pwr.edu.pl}
\affiliation{Department of Experimental Physics, Faculty of Fundamental Problems of Technology, Wroc\l{}aw University of Science and Technology, Wybrzeże Wyspia\'{n}skiego 27, 50-370 Wroc\l{}aw, Poland}

\author{M.~Pieczarka}
\affiliation{Department of Experimental Physics, Faculty of Fundamental Problems of Technology, Wroc\l{}aw University of Science and Technology, Wybrzeże Wyspia\'{n}skiego 27, 50-370 Wroc\l{}aw, Poland}

\author{C.~Schneider}
\affiliation{Carl von Ossietzky Universit\"{a}t Oldenburg, Fakult\"{a}t V, Institut f\"{u}r Physik, 26129 Oldenburg, Germany}

\author{S.~H\"{o}fling}
\affiliation{Julius-Maximilians-Universit\"{a}t W\"{u}rzburg, Physikalisches Institut and W\"{u}rzburg-Dresden Cluster of Excellence ct.qmat, Lehrstuhl f\"{u}r Technische Physik, Am Hubland, 97074 W\"{u}rzburg, Germany}

\author{S.~Klembt}
\affiliation{Julius-Maximilians-Universit\"{a}t W\"{u}rzburg, Physikalisches Institut and W\"{u}rzburg-Dresden Cluster of Excellence ct.qmat, Lehrstuhl f\"{u}r Technische Physik, Am Hubland, 97074 W\"{u}rzburg, Germany}

\author{M.~Syperek}
\affiliation{Department of Experimental Physics, Faculty of Fundamental Problems of Technology, Wroc\l{}aw University of Science and Technology, Wybrzeże Wyspia\'{n}skiego 27, 50-370 Wroc\l{}aw, Poland}

\begin{abstract}


Although energy level repulsion is typically observed in interacting quantum systems, non-Hermitian physics predicts the effect of level attraction, which occurs when significant energy dissipation is present. Here, we show a manifestation of dissipative coupling in a high-quality AlGaAs-based polariton microcavity, where two polariton branches attract, resulting in an anomalous, inverted dispersion of the lower branch in momentum dispersion. Using angle-resolved photoluminescence measurements we observe the evolution of the level attraction with exciton-photon detuning, leading to changes in anomalous dispersion shape within a single sample. The dissipative coupling is explained by the interaction with an indirect exciton, acting as a highly dissipative channel in our system, and the observed dispersions are well captured within a phenomenological model. Our results present a new mechanism of dissipative coupling in light-matter systems and offer a tunable and well-controlled AlGaAs-based platform for engineering the non-Hermitian and negative mass effects in polariton systems.

\end{abstract}

\maketitle

\section*{Introduction}

\begin{figure*}
    \centering
    \includegraphics[width=13 cm]{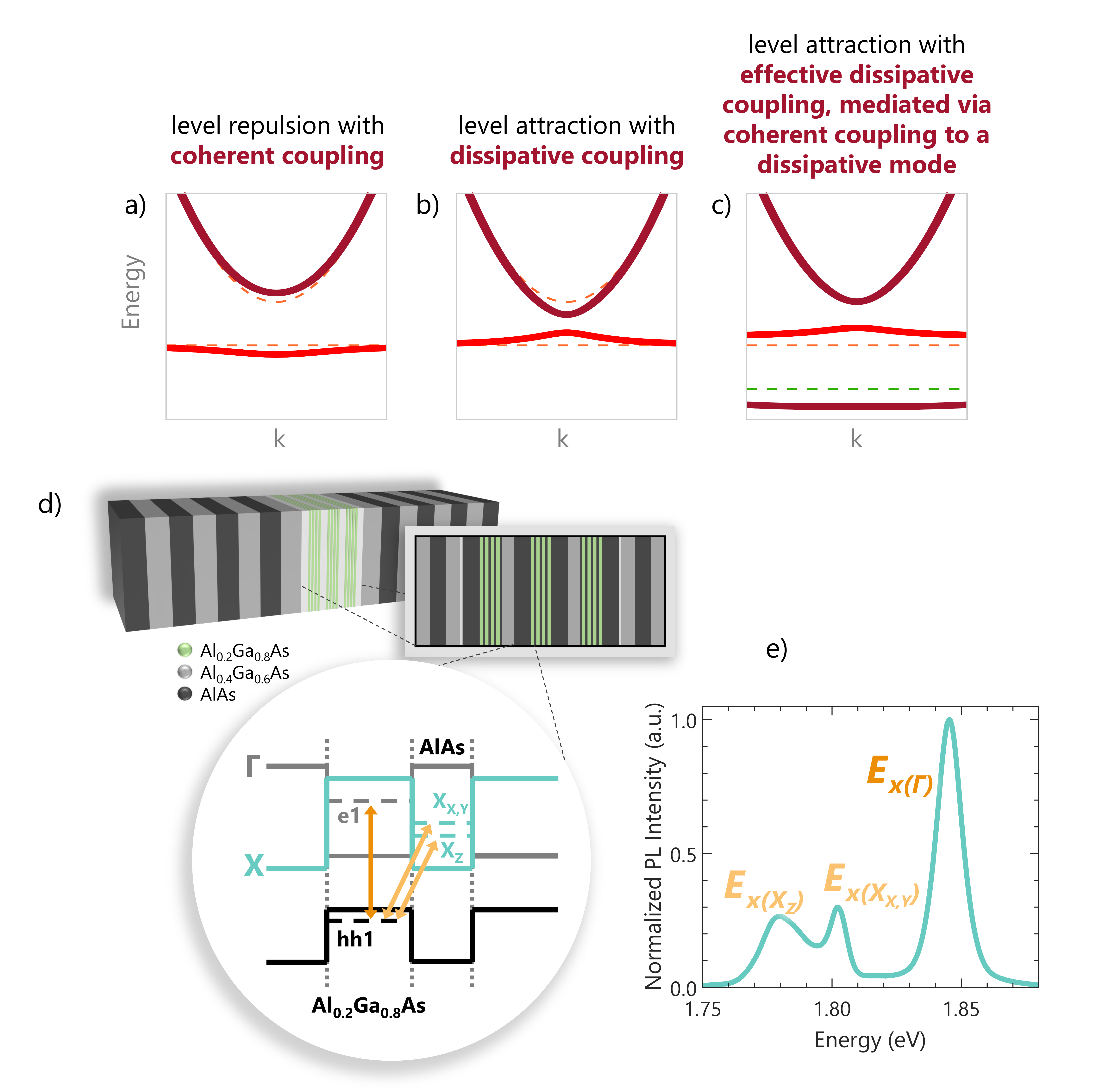}
    \caption{A schematic visualisation of the level attraction effect in an energy-momentum dispersion (a-c) and the investigated structure (d-e). (a) Levels of a strongly coupled system of two coherently coupled modes, showing level repulsion. (b) Level attraction of the same two modes, coupled with an imaginary coupling. (c) A schematic visualisation of a similar level attraction effect, but coming from a real coupling between three modes, one of which is strongly dissipative. Two initial modes in (a-c) are marked with dashed lines, and the third, dissipative mode in (c) is highlighted in green. (d) Schematics of the investigated microcavity, with a close-up of the active layer. In the system band structure solid lines show the edges of the X, ${\Gamma}$ and valence bands of one period of the repeated layers. Dashed lines indicate the quantized electron ($e1$, $X_{X,Y}$, $X_Z$) and heavy hole ($hh1$) levels in two adjacent layers. Carriers occupying these levels subsequently form three excitons present within the system (indicated with orange arrows), when subject to Coulomb interactions. (e) The photoluminescence spectrum of the bare quantum well system, with the top Bragg reflector etched away. Three well-resolved features, labeled as $E_{x(\Gamma)}$, $E_{x(X_{X,Y})}$ and $E_{x(X_Z)}$, correspond to transitions of three excitonic species present in our sample.}
    \label{fig1}
\end{figure*}

In interacting quantum systems, it is typical to observe level repulsion.  When two modes couple and intermix, the resulting energy levels anticross, avoiding degeneracy at resonance.  If the strongly interacting states are photons and excitons confined in planar microcavities, the resulting eigenstates appear as two exciton-polariton branches, schematically depicted in Fig. \ref{fig1}(a). Lower polaritons are characterized by a nearly parabolic dispersion at small wavevectors, with small and positive effective mass, inherited largely from the photonic component. At larger momenta characteristic inflection points appear, around which the second derivative of the energy dispersion changes sign, nevertheless, the mass that determines the group velocity remains positive for all momenta \cite{Colas2018}. 
The mode dispersion and particle effective mass can be further engineered, typically by introducing an additional potential landscape in the system, such as lattice potentials, yet it requires additional sample processing or sophisticated excitation schemes \cite{Schneider_2017,Tanese2013,Whittaker2018,Gianfrate2024}. 

However, in all open systems, losses are inevitable, and the interactions and eigenstates are strongly affected by dissipation. Optical systems in which light confinement can be effectively engineered, such as high-quality optical microcavities, are an ideal experimental platform to study dissipation-related coupling effects. When dissipation becomes equally important to the coherent coupling, the emergent states can attract (instead of repelling), even without additional potential. The attraction effect is analogous to classical in-phase oscillations of dissipatively coupled pendulums \cite{HarderJAP2021}. In light-matter systems the influence of dissipative coupling has been experimentally observed in photonic-crystal cavities containing single quantum dots \cite{Tawara2010}. In two-dimensional polaritonic systems however, while there were some first experimental hints in rather low-Q microcavities containing monolayer semiconductors \cite{Dhara2018,Wurdack2023}, clear studies in narrow-linewidth systems are elusive so far. The level attraction phenomenon has mainly been studied in 
other contexts, such as magnons \cite{HarderPRL2018,HarderJAP2021,Wang2020}, microwave cavities \cite{Grigoryan2018}, opto-mechano-fluidic resonators \cite{Lu2023} or mechanical systems \cite{Elste2009}. Dissipative coupling has been suggested as a potential mechanism for entangled state creation, as a new tool in the design of superconducting qubits \cite{HarderPRL2018,Wang2020}, for development of metamaterials \cite{HarderPRL2018}, but also as a mechanism beneficial in cavity spintronics \cite{Wang2020}. 

When the coupling of two quantum mechanical oscillators with parabolic dispersions, subject to substantial loss, becomes complex, and the imaginary (dissipative) coupling is comparable to the real (coherent) coupling, the levels attract, and the dispersion of one of the modes can invert, presenting an anomalous behaviour. It is visualized in Fig. \ref{fig1}(b). The resulted band has a negative curvature parabolic wavevector dependence, directly representing the negative effective mass of the emergent state. 
However, even though the level attraction and the anomalous dispersion can be phenomenologically described by the imaginary coupling between the two states, the physical origin of the effect is not always clear and varies between systems. Interestingly, it has been shown how, in a physical system, the dissipative type of coupling can be realized by coupling two oscillators reactively to a third, highly dissipative entity \cite{YuPRL2019}, as schematically depicted in Fig. \ref{fig1}(c). The third-party mode in cavity systems can come from an invisible cavity mode with extremely high leakage or dissipation. This mechanism has been successfully used to explain the level attraction in magnon cavities \cite{YuPRL2019}, yet has never been considered in an exciton-polariton context. 

Regardless of the mechanism, the negative mass of such an inverted state can be used in a wide range of studies on non-Hermitian effects or topology \cite{Parto2021,Gao2015,long2022}. It manifests itself in the particle’s dynamics, so that its group velocity and momentum have opposing directions \cite{Wurdack2023,Colas2018}. Next to substantial fundamental interest, this, in turn, can be employed to control wavepacket dynamics \cite{Colas2018}, hydrodynamics \cite{Ballarini2017}, or cause resonance trapping \cite{Persson2000}. For all these applications, engineering the inverted dispersion is crucial, yet so far cavity engineering  focused mainly on the potential engineering or spin-orbit interactions in polariton microcavities, rather than the dissipation. Precise control over the attraction strength would also be hugely beneficial. 

In exciton-polariton settings, anomalous dispersion has been predicted \cite{Bleu2024}, but it has been experimentally observed only very recently and only in transition-metal dichalcogenide samples \cite{Dhara2018,Wurdack2023}. This medium lacks the exciton energy control and ease of the cavity design of a III-V based semiconductor and proved to be challenging in reproducibility. Moreover, in the experimental hints made so far, the effect was strongly obscured by inhomogeneously broadened lines, while their theoretical descriptions vary widely.


In this work, we unequivocally demonstrate the level attraction manifested as an inverted anomalous dispersion in the AlGaAs exciton-polariton system. We investigate the mechanism of dissipation in our structure, crucial for the attraction to occur. In contrast to previous studies, our III-V semiconductor sample not only hosts conventionally studied $\Gamma$-excitons in the QWs, coherently coupled to photons, but also lower-energy spatially- and momentum-indirect X-excitons, which are strongly prone to dissipation. We show that the source of dissipation in our structure is the lower-energy indirect state, acting as a draining channel for both photons and electrons. This highly dissipative mode allows for the dissipative coupling to become sufficiently strong to surpass the coherent exciton-photon coupling, and result in inverted eigenstate dispersion. Finally, we demonstrate the superiority of our material system in comparison to previous realisations, owing to its high tunability, ease of design, and huge potential for non-Hermitian phases engineering, by showing a change of the dispersion shape as a function of exciton-photon detuning.

\section*{Results}
\subsection*{Excitonic structure}

We studied an AlGaAs/AlAs optical microcavity, designed for room temperature polaritonics \cite{SuchomelOE2017}. However, in this work we focus on experimental observations made at cryogenic temperature of 4 K, benefiting from the narrow polaritonic linewidths. The sample schematic is depicted in Fig.~\ref{fig1}(d) and a detailed description of the sample composition can be found in Methods.

Due to the high aluminium content affecting the band alignment, the structure hosts direct and indirect excitons in the quantum well (QW) \cite{Chand1984,SuchomelOE2017,BieganskaARXIV}. Apart from the conventional direct excitons composed of $\Gamma$-valley electrons and heavy holes confined in the QW layer, the structure also hosts lower-energy spatially and momentum indirect X-excitons \cite{BieganskaARXIV}. Since the order of X and $\Gamma$-valley energy minimum in the conduction band is reversed for the Al$_{0.2}$Ga$_{0.8}$As QW and for the AlAs barrier material, the fundamental QW electron state resides in the barrier. This
allows the formation of indirect excitons composed of X-valley electrons in the barrier Coulomb-correlated with $\Gamma$-valley heavy holes confined in the QW layer. Two lowest-energy optically active states relate to excitons consisting of X-valley electrons with different effective masses (longitudinal and transverse with respect to the spatial quantization axis), forming $X_Z$ and $X_{X,Y}$ states respectively. The single-particle energy levels are visualized in the QW band structure in Fig. \ref{fig1}(d), using dashed lines. Measured spectrum of the bare QW active material is presented in Fig. \ref{fig1}(e), where all excitonic transitions are indicated. The indirect nature of these excitonic states has been investigated in detail in our previous work \cite{BieganskaARXIV}. 

When embedded in a monolithic optical microcavity close to resonance with the $\Gamma$-state, direct excitons couple strongly to light, forming exciton-polariton quasiparticles \cite{SuchomelOE2017}. These states are characterized by the normal-mode splitting and present typical polariton dispersions (as shown in Supplementary Material, section I). However, herein we study the structure at very large negative $\Gamma$-exciton – photon detunings, $\Delta_{\Gamma}=E_c-E_{x(\Gamma)} <0$ (where $E_c$ is the cavity mode energy and $E_{x(\Gamma)}$  is the energy of the direct exciton in the QW). In this regime, the light-matter interactions are dominated by the coupling of the cavity optical mode to the indirect X -valley excitons and the resulting states strongly differ from the typical exciton-polaritons under coherent light-matter coupling. The detuning is sufficiently large that the coherent coupling to the $\Gamma$-excitons becomes irrelevant. For convenience, throughout the rest of the paper, we will refer to the detuning as defined with respect to the higher-energy X-exciton, $\Delta_X = E_c-E_{x(X_{X,Y})}$.


\subsection*{Photoluminescence Measurements}

\begin{figure*}[ht]
    \centering
    \includegraphics[width=\textwidth]{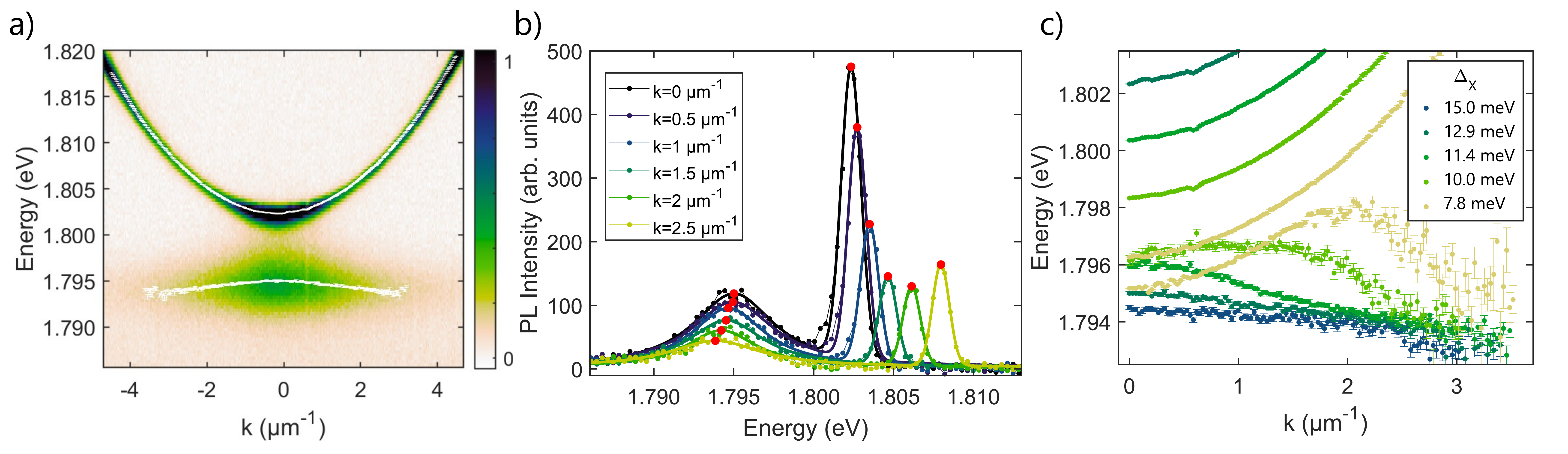}
    \caption{Experimental observation of the anomalous dispersion. (a)  Momentum-resolved photoluminescence image at a chosen exciton-photon detuning (linear color scale). Spectra crossections taken at several wavevectors are presented in (b) (connected dots), together with fitted curves (solid lines). Red dots show the energies of the two deconvoluted modes, extracted from fitting, also marked in white in (a). (c) Extracted mode dispersions at several exciton-photon detunings $\Delta_X$. Error bars indicate the fitting standard error.}
    \label{fig2}
\end{figure*}

To study the coupling between photons and X-excitons, we measured angle-resolved photoluminescence spectra in a wide detuning range, close to resonance with the X-excitons. When the photonic mode gets sufficiently close to the energy of the $X_{X,Y}$ excitonic resonance, a new lower energy state brightens up, with the dispersion curved in a distinctly inverted manner. An experimental example of such a momentum dispersion is presented in Fig.~\ref{fig2}(a), together with the peak energies of the two branches, extracted with a fitting procedure (see Methods and Supplementary Material, section II). An apparent and monotonous redshift of this mode’s energy with increasing wavevector can be seen in Fig.~\ref{fig2}(b), a dependence opposite to the higher energy photonic-like state. The two levels attract, causing the mirroring of their wavevector energy dispersions, mimicking the dispersion sketched in Fig.~\ref{fig1}(b). The negative curvature of such an inverted parabolic dispersion is directly linked to the negative effective mass of the lower mode - a rare phenomenon in exciton-polariton systems \cite{Colas2016,Wurdack2023,Dhara2018,Paik2023}. 

Taking advantage of the cavity energy gradient (due to the thickness variation across the sample), we probed the negative mass states in a range of sample positions (detunings). As presented in Fig.~\ref{fig2}(c), decreasing the detuning between the cavity mode and the $X_{X,Y}$-exciton energy $\Delta_X$ leads to an increase in attraction effect, with the anomalous shape of the lower branch becoming steeper and more distinct. Figure \ref{fig2}(c) shows the energies of two polaritonic branches extracted from fitting the PL measurements taken at different sample positions. Interestingly, around the positive photon to $X_{X,Y}$-exciton detuning of approximately $10\ meV$ the curvature changes from the inverted parabola-like with one energy maximum at $k=0$ to anomalous shape with two distinct and symmetric maxima at $k\neq0$. Similar dispersion shapes have been observed before in different structures in both regimes \cite{Dhara2018,Wurdack2023}, yet never in the same material system, nor in a single sample. The corresponding change of the effective mass value with the detuning is presented in Supplementary Material, section VI.

At negative detunings $\Delta_X$, only one branch appears in the photoluminescence spectrum, with the standard parabolic shape of the dispersion resembling the one of a photonic mode, as presented in Supplementary Material, section I. For further discussions, we focus solely on the level attraction region.


\subsection*{Model}

\begin{figure*}[ht]
    \centering
    \includegraphics[width=\textwidth]{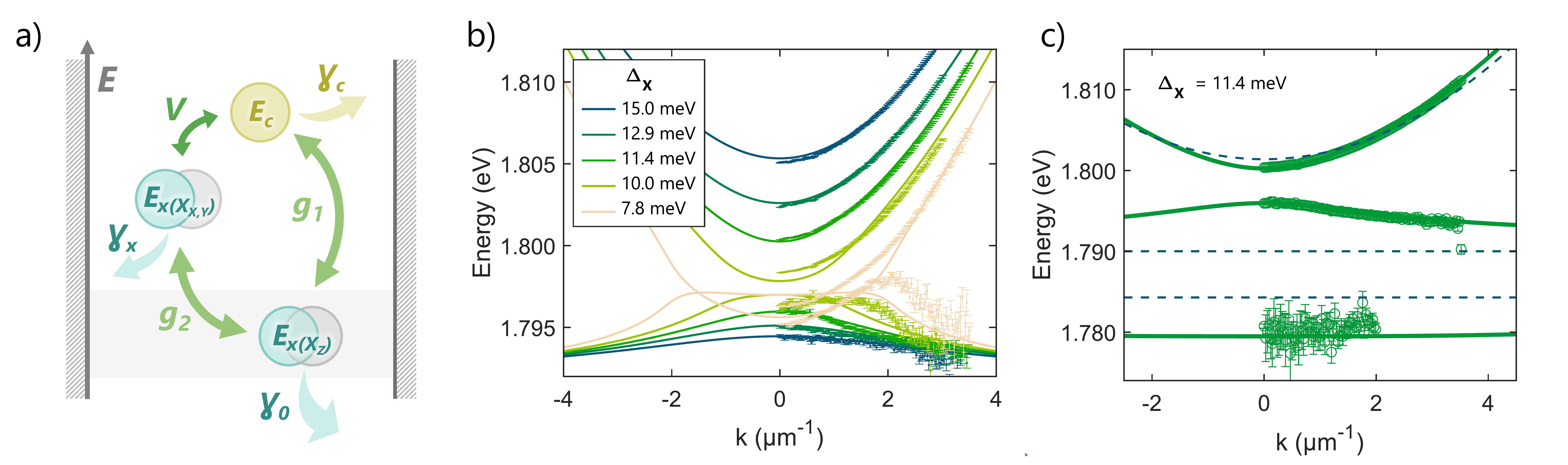}
    \caption{Level attraction modelling. (a) Schematic visualisation of the three coupled oscillators model and its application in our system. The coupled particles (photon ($E_c$, presented as a single circle) and two indirect excitons ($E_{x(X_{X,Y})}$ and $E_{x(X_Z)}$), represented as double circles) – are shown on a schematic energy scale, with their intrinsic decays sketched as broad arrows, while the couplings are presented as two-sided arrows. (b) Comparison of the model lines with experimental level branches at several exciton-photon detunings, plotted in corresponding colors. Model parameters are described in the main text. (c)  Example dispersion at a single photon-$X_{X,Y}$-exciton detuning of $11.4\ meV$. Dashed lines mark the dispersions of a bare photonic mode and two indirect excitons, open points are the fitted peak positions of the three polaritonic branches, and solid green lines are the model dispersions. For clarity, the experimental data is only presented for positive k.}
    \label{fig3}
\end{figure*}

To understand the source of level attraction, we have to recall the existence of the $X_Z$-electron exciton, with energy below both the $X_{X,Y}$ exciton and the photonic mode, which inclusion proves to be crucial in the theoretical description of the data. To describe our system and quantify the mechanism of level attraction, we used a general three coupled oscillator model, predicting the attractive level crossing via the existence of a dissipative mode \cite{YuPRL2019}. Even though the level attraction has been previously described with the use of imaginary coupling between two oscillators \cite{Wurdack2023,Bleu2024}, in \cite{YuPRL2019} the authors show how, in a physical system, the dissipative coupling can be realized by coupling two oscillators to a third highly dissipative one, even if the mode is invisible. In our case, the third-party mode could be identified as the $X_Z$ -exciton.

The model can be represented by a $3 \times 3$ non-Hermitian matrix:

\begin{equation*}
\begin{split}
& H=\left(\begin{matrix}E_1&V&g_1\\V&E_2&g_2\\g_1&g_2&E_0\\\end{matrix}\right)\\
 & =\left(\begin{matrix}E_c-i\gamma_c&V&g_1\\V&E_{x\left(X_{X,Y}\right)}-i\gamma_x&g_2\\g_1&g_2&E_{x\left(X_Z\right)}-i\gamma_0\\\end{matrix}\right).
\end{split}
\end{equation*}

In this approach, two oscillators with intrinsic decay (with energies of $E_1$ and $E_2$) are coupled to each other coherently via $V$, and to the third oscillator $E_0$, which is strongly damped, $\gamma_0 \gg \gamma_c, \gamma_x$. Significant dissipation of the third state is crucial for the level attraction and for $E_0$’s strong influence on the $E_1$ and $E_2$ dispersions, when the real coupling terms $g_1$ and $g_2$ are sufficiently large to surpass the coherent coupling $V$. In such conditions, these terms can effectively act as complex coupling between the two modes \cite{YuPRL2019,HarderPRL2018,Wurdack2023}, provided that $E_1$ and $E_2$ are nearly resonant, what is further discussed in Supplementary Material, section III. In a regime of high coherent coupling between the two resonances and a weak dissipation of the third mode all eigenstates repel, as it is typically observed in exciton-polariton systems \cite{Kavokin_microcavities2008,Wurdack2017,Koksal2021,Zhumagulov2022,Hofner2015}.

We schematically visualise the model and the involved oscillators in Fig.~\ref{fig3}(a). In our structure, two coupled resonances are the photonic mode $C$ and the $X_{X,Y}$ exciton, with energies and decay rates of $E_c$, $E_{x(X_{X,Y})}$  and  $\gamma_c$, $\gamma_x$  respectively. The lower-energy $X_Z$ excitonic resonance acts as a dissipative mode and is characterized by the energy of $E_{x(X_Z)}$  and dissipation $\gamma_0$. The coupling between photons and $X_{X,Y}$ excitons inside the microcavity ($V$) is expected to be weak, due to the space- and momentum- indirect nature of the excitonic resonance. On the contrary, $X_Z$-exciton is expected to couple to light more efficiently, as the spatial symmetry breaking allows for its recombination without the assistance of phonons, due to the weakening of the momentum-conservation rules, regardless of its indirect nature \cite{Pietka2007,Danan1987,BieganskaARXIV}. The coupling between the two indirect excitons is enabled via transfer of electrons between the states and transitions from the higher  $X_{X,Y}$ to the lower $X_Z$ electronic state, as evidenced by complex temporal dynamics \cite{BieganskaARXIV} and previous studies \cite{Feldmann1992,Finkman1987,Wysmoek2004,Peter1989}. Both couplings $g_1$ and $g_2$ are therefore expected to play a significant role in the system, with the $g_1$ value expected to be much larger than $V$. The energies of both excitonic resonances can be directly inferred from the photoluminescence measurements of the bare QW structure (see Fig.~\ref{fig1}(e) and \cite{BieganskaARXIV}). 

Using this approach, we modelled our experimental dispersions
, as presented in Fig. \ref{fig3}(b). Experimental points are the extracted peak energies of the two polaritonic branches at several exciton-photon detunings $\Delta_X$, and solid lines show the fitted model eigenstates. Additionally, in Fig.~\ref{fig3}(c) we show all three of the model eigenstates at the exciton-photon detuning of $11.4 ~meV$, as well as the dispersions of a bare photonic mode and two indirect excitons (dashed lines). 
We note that in most measurements, the lowest-energy mode cannot be seen in the photoluminescence spectra, except near the $\Delta_X \approx 11\ meV$ detuning, hence we used only two states in the dispersion modelling.

Model results show very good correspondence with the measured dispersions. The model reflects well the anomalous shape of the lower branch dispersion and captures a clear transition between its monotonic (with a single maximum at $k=0$) and non-monotonic (with maxima at finite wavevectors) $|k|$-dependence when decreasing $\Delta_X$. At larger detunings the model dispersions match experimental points nearly perfectly, demonstrating the change in curvature around $k=0$, linked to the dissipative level attraction. Discrepancies between the model and the experimental curves become visible only at smaller positive exciton-photon detunings $(\Delta_X\le10\ meV)$. 
This may arise from the fact, that to model our data we set all the parameters constant throughout this detuning range (apart from the photonic mode energy), which is a simplified approach. 
All three decay constants, as well as level energies, can vary across the sample, due to the local disorder and the layer width change. Nevertheless, the model describes our system very well in a large range of exciton-photon detunings, even when using only one set of parameters. Moreover, a high agreement between the model line and the third state detectable at the detuning $\Delta_X = 11.4\ meV$ presented in Fig.~\ref{fig3}(c), despite not using this state in the fitting, further proves the applicability of our model.

The extracted exciton-photon couplings are $V=0.1\ meV$ and $g_1=10.6\ meV$, while the coupling between two X-excitons $g_2$ is $17\ meV$. As expected, the coherent coupling between the photonic mode and the spatially and momentum indirect $X_{X,Y}$ exciton is much smaller than other energies in our system. The highly dissipative $X_Z$ state couples to light more efficiently, what is likely a result of the symmetry breaking effect described above. The most influential interaction comes from the nonradiative coupling between the two X-excitons. The extracted decay rates of all states are $\gamma_c=0.1\ meV$, $\gamma_x=0.01\ meV$ and $\gamma_0=41\ meV$. The model photon linewidth value corresponds to a lifetime of approximately $\sim 6\ ps$, which is a value expected for this microcavity, subject to disorder and operating 
at large detuning from the designed wavelength \cite{Panzarini1999}. A small line broadening of the $X_{X,Y}$ state originates from its longer lifetime, expected from its indirect nature. On the other hand, large broadening $\gamma_0$ of the $X_Z$ exciton points to its dissipative role and it is crucial to obtain level attraction in our system. We note that the model value is larger than the measured photoluminescence linewidth broadening of this state of $\sim 20\ meV$, measured with the top mirror removed from the cavity \cite{BieganskaARXIV}. However, the observed emission linewidth cannot be directly translated into the homogeneous broadening. Photoluminescence broadening consists of both homogeneous and inhomogeneous parts, but, at the same time, can be narrowed by a Purcell effect, resulting from a formation of very low-Q-factor half-microcavity \cite{Kavokin_microcavities2008}. Large damping of this mode likely comes from the sensitivity of these states to structure inhomogeneities, stemming from their ground state nature, and affecting their lifetime and transport properties, as discussed in detail in \cite{BieganskaARXIV} and shown before \cite{Lee1996}. Overall, the model accurately describes our system and reveals the highly damped $X_Z$  excitons as the source of the level attraction and the inverted polariton dispersion. The importance of the damped mode inclusion is presented in section IV of the Supplementary Material.

In addition, we considered the contribution of the three involved oscillators in the final system eigenvalues, by studying the Hopfield coefficients \cite{Vasilevskiy2015,Hopfield1958}. Coefficient wavevector dependencies reflect the anomalous behaviour of the inverted anomalous branch, with the dissipative exciton fraction gaining importance in the anomalous region (at small wavevectors), particularly at small detunings. It further highlights the importance of the $X_Z$ excitons in the observed effect. Hopfield dispersions at several detunings, as well as their more detailed discussion, can be found in section V of the Supplementary Material.


\section*{Discussion}

In summary, we have observed the anomalous dispersion of the polaritonic branch in an AlGaAs-based microcavity, characterized by the negative effective mass. Our AlGaAs-based semiconductor system offers precise high-quality growth and design of the layers, fine-tuning its properties, what will uniquely allow to tailor the coherent coupling and the optical Q-factor as well as the dissipation, by engineering the $X_{X,Y}$ and $X_Z$ excitons. 
We have shown how the presence of and the coupling to the indirect excitonic state energetically below the excitonic and photonic resonances, which acts as a channel of loss, can manifest itself as a dissipative coupling between these states. Our hypothesis is supported by a phenomenological model of three coupled oscillators. The high dissipation rate of this indirect state is crucial to make the effective coupling non-Hermitian and overcome the coherent coupling. Furthermore, we have observed the evolution of the system eigenstates with varying detuning, showing the shift and the change of the eigenstate dispersion curvature. We show two regimes of anomalous dispersion shape, with eigenstate energy maxima at $k=0$ and $k\neq 0$ in a single sample.

Previously, the anomalous dispersion of exciton polaritons in unstructured samples has been observed solely in transition metal dichalcogenide-based samples \cite{Dhara2018,Wurdack2023,Paik2023}. The supporting models were applicable only for many-particle excitons in heavily-doped samples with complex interactions \cite{Dhara2018}, or for media with a strong influence of exciton-phonon interactions \cite{Wurdack2023}. In our case of a III-V semiconductor-based system at cryogenic temperature, phonon influence is known to be much smaller, hence insufficient to lead to the dissipative coupling. Previously studied systems lacked the presence of a tunable and energetically-lower state providing a channel of loss, which proves to be crucial in our structure. Even more importantly, they also lacked the excellent linewidths, making the dispersion shape less distinct and rendering interpretations of the observed dispersions less robust.  

Here, we present the effect of level attraction in a new experimental platform, with a new source of the dissipative coupling in the exciton-polariton context. Apart from narrow linewidths, our system provides a great opportunity for level attraction tuning, via changing the detuning between the resonances, owing to the sample's wedged growth. So far, the effective mass engineering had to involve additional sample processing steps (such as patterning or etching) or complicated excitation schemes, e.g. structured beams, which prove to be costly, imperfect, and often difficult to implement. In our case the mass can be engineered during the typical sample growth, with no further steps required. The change of the eigenstate dispersion curvature can be easily accessed by simply changing the position on the sample, allowing access to different anomalous dispersion curvatures in a single platform. Such a straightforward tuning was lacking in previous observations, and shows a clear path for future studies or device design. 

Anomalous dispersion can be employed in novel studies of non-Hermitian effects \cite{Parto2021,Feng2017,Gao2015}, nontrivial dynamics and hydrodynamics \cite{Colas2018,Khamehchi2017} and in studies of analogue systems \cite{Pickup2020,Jacquet2020,StJean2021}. It allows access to a plethora of studies on the exceptional points and related phenomena, such as winding of the complex eigenenergies, chiral modes, topological lasing, or enhanced perturbation, among others \cite{Su2021,Baboux2018,long2022,Ardizzone2022}. The optical, easily experimentally-accessible platform of high quality and high adjustability presented in this work makes our finding relevant and desirable far beyond the exciton-polariton context. The observation can be considered to be an important contribution to the broad field of dissipative coupling effects, with losses and dissipation affecting practically all physical systems.


\section*{Methods}

\subsection*{Sample}
The sample under study consists of twelve 9 nm-wide Al$_{0.20}$Ga$_{0.80}$As QWs, separated by 4 nm AlAs barriers, distributed in three stacks of four (as visualized in Fig. \ref{fig1}(d)). The stacks are placed in a $\lambda$/2-AlAs cavity surrounded by AlAs/Al$_{0.40}$Ga$_{0.60}$As distributed Bragg reflectors (DBRs), consisting of 28/24 mirror pairs in the bottom/top reflector, including 3 nm GaAs smoothing layers after each mirror pair in the local minimum of the electromagnetic field. The whole microcavity structure was grown by molecular beam epitaxy on the GaAs substrate. Lack of wafer rotation during growth results in a gradual change of the cavity length across the sample, allowing for the experimental access to a wide range of exciton-photon detunings. The photoluminescence spectrum of the bare quantum well system presented in Fig.~\ref{fig1}(e) was taken on a sample with the
top Bragg reflector etched away \cite{BieganskaARXIV}.
\\

\subsection*{Optical Measurements}
The sample was placed in the continuous flow liquid helium cryostat and cooled down to 4.2 K. It was excited by laser pulses from the OPO pumped by a Ti:Sapphire pulsed laser with 76 MHz repetition rate, generating the wavelength of around 620 nm. The beam was focused on a sample via a NA = 0.65 objective. Structure photoluminescence was then collected by the same objective and imaged on a slit of a monochromator (with a 1200 lines/cm groove density diffraction grating) equipped with a high-efficiency EMCCD camera. Imaging the Fourier plane by using four confocal lenses in the detection path allowed for the angle-resolved measurements.
\\

\subsection*{Dispersion extraction}
Photoluminescence spectra at each wavevector were fitted with a sum of a Lorentzian (lower central energy) and Gaussian (higher energy) curves. The extracted peak energies were used for further modelling. The error bars presented throughout the manuscript come from the fitting standard error. An exemplary fit with the more detailed discussion on the fitting function selection can be found in the Supplementary Material, section II.


\section*{Data availability}
The datasets generated during and/or analysed during the current study are available from the corresponding author on reasonable request.

\bibliography{bibliography} 

\section*{Acknowledgements}
D. B., M.P. and M. S. acknowledge financial support from the National Science Centre Poland within the Grant No. 2018/30/E/ST7/00648. 
C.S. and S.K. gratefully acknowledge funding by the German Research Association (DFG) within the project SCHN1376 13.1 and KL 3124/3-1 (El Pollo Loco).
S.H. and S.K. acknowledge financial support by the DFG under Germany’s Excellence Strategy - EXC2147 ct.qmat (Project No. 390 858 490).
S.H. acknowledges financial support by the DFG project HO 5194/12-1.

\section*{Author contributions statement}
D. B. conducted all the spectroscopic experiments and analysed the experimental data. D.B. and M.P. performed theoretical modeling of the results. S. K., S. H, and C. S provided the sample. D.B., M.P., M.S. analyzed the results and all authors contributed to their discussion. D.B. wrote the first version of the manuscript and prepared all figures. All authors reviewed the manuscript to its final form.

\section*{Additional information}
The authors declare no competing interests. Supplementary Information is available for this paper.

\end{document}